\documentclass[num-refs]{wiley-article}

\usepackage{siunitx}

\usepackage{epstopdf}

\papertype{Original Article}
\paperfield{}

\title{Capillary adhesion of stick insects} 

\abbrevs{FTIR, frustrated total internal reflection.}

\author[1]{Guillermo J. Amador}
\author[1]{Brett Klaassen van Oorschot}
\author[2]{Uddalok Sen}
\author[3]{Benjamin Karman}
\author[1]{Rutger Leenders}

\affil[1]{Experimental Zoology group, Wageningen University \& Research, Wageningen, 6708 WD, The Netherlands}
\affil[2]{Physical Chemistry and Soft Matter, Wageningen University \& Research, Wageningen, 6708 WE, The Netherlands}
\affil[3]{Biology Department, Vrije Universiteit Amsterdam, Amsterdam, 1081 HV, The Netherlands}

\corraddress{Guillermo J. Amador, Experimental Zoology group, Wageningen University \& Research, Wageningen, 6708 WD, The Netherlands}
\corremail{guillermo.amador@wur.nl}

\fundinginfo{Netherlands Organisation for Scientific Research (NWO), Science Domain, VI.Vidi.213.122; Wageningen Institute of Animal Sciences (WIAS), Postdoc Talent Programme}

\runningauthor{Amador et al.}

\begin{document}

\begin{frontmatter}
\maketitle

\begin{abstract}
Scientific progress within the last few decades has revealed the functional morphology of an insect’s sticky footpads---a soft, sponge-like pad that secretes a thin liquid film. However, the physico-chemical mechanisms underlying their adhesion remain elusive. Here, we explore these underlying mechanisms by simultaneously measuring adhesive force and contact geometry of the adhesive footpads of live, tethered Indian stick insects, \textit{Carausius morosus}, spanning more than two orders of magnitude in body mass. We find that the adhesive force we measure is similar to previous measurements that use a centrifuge. Our measurements afford us the opportunity to directly probe the adhesive stress \textit{in vivo}, and use existing theory on capillary adhesion to predict the surface tension of the secreted liquid and compare it to previous assumptions. From our predictions, we find that the surface tension required to generate the adhesive stresses we observed ranges between 0.68 mN m$^{-1}$ and 12 mN m$^{-1}$. The low surface tension of the liquid would enhance the wetting of the stick insect's footpads and promote their ability to conform to various substrates. Our insights may inform the biomimetic design of capillary-based, reversible adhesives and motivate future studies on the capillary properties of the secreted liquid.


\end{abstract}
\end{frontmatter}

\section*{Introduction}
Nature is often the source of inspiration for developing new technologies, leading to the growth of the field of biomimetics \cite{bhushan-2009-philtransrsoca}, and adhesion is not an exception. Adhesion is the ability of a substance to stick (or adhere) to a dissimilar substance. Some species of insects can generate enormous adhesive forces. For example, Asian weaver ants (\textit{Oecophylla smaragdina}) have been observed to adhere to glass substrates upside down while supporting loads more than 100 times their own body weight \cite{dirks-2011-softmatter}. Over the course of their lifetime, insects adhere to a multitude of substrates of varying roughness and wettabilities. The versatility of insect adhesion mechanisms are promising for applications involving bioinspired adhesives \cite{feldmann-2021-advmaterinterfaces}, particularly in the design of robotic manipulators and climbers \cite{drotlef-2019-integrcompbiol}.

Any adhesive should primarily satisfy two requirements: (i) establish a good contact with the substrate, even in the presence of roughness, and (ii) dissipate a significant amount of energy during separation \cite{gay-1999-phystoday}. However, it is expected that modern (and future) adhesives would do more than just stick \cite{fakley-2001-chemind}, leading to a surge in recent years towards understanding and developing newer adhesion mechanisms. The advancement of microscopes along with development of superior analytical tools have led to a renewed interest among biologists and engineers towards the development of biomimetic adhesives \cite{book-smith}. Nonetheless, before one delves deeper into the engineering details of biomimetic adhesion, there are several fundamental questions that need answers \cite{dirks-2011-softmatter}. Although high-precision characterization techniques such as atomic force microscopy (AFM) and scanning electron microscopy (SEM) have enabled biologists to examine the topology of the insect footpads down to the nanometer-scale, the detailed mechanisms of biological adhesion are still not fully understood \cite{federle-2006-jexpbiol}, particularly what the underlying physics and chemistry are and how to represent them in a mathematical model.

\begin{figure}[bt]
\centering
\includegraphics[width=\textwidth]{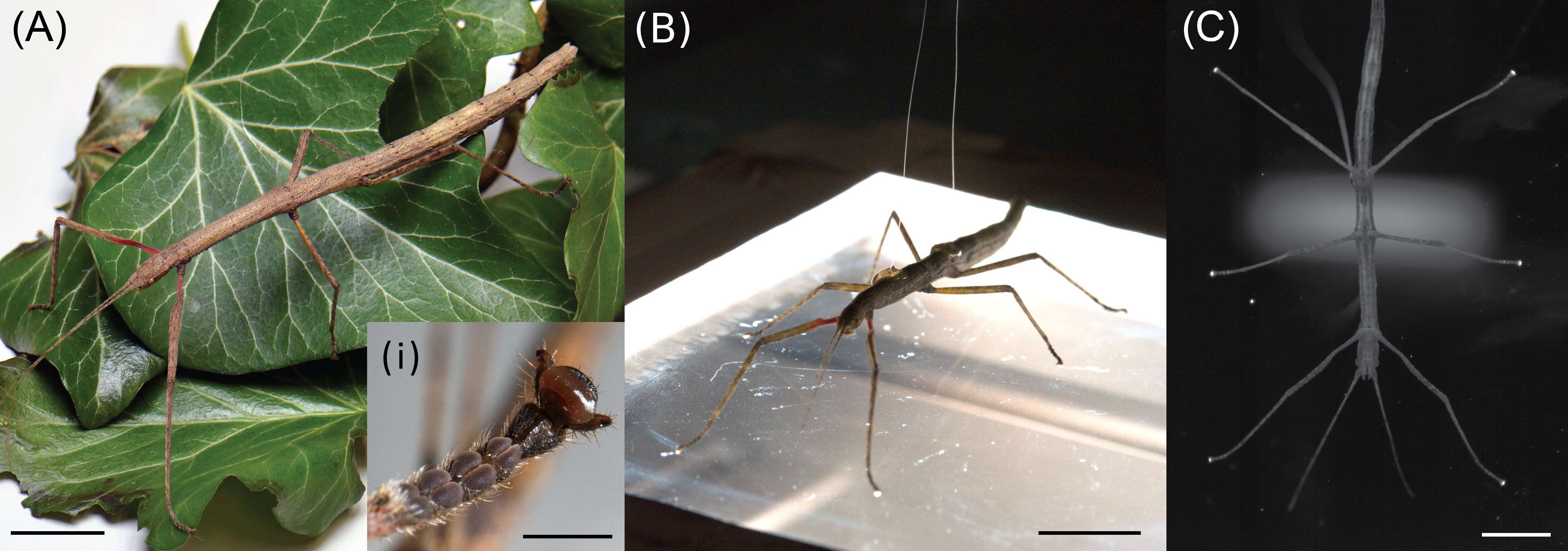}
\caption{(A) An Indian stick insect (\textit{Carausius morosus}) and (i) its distal tarsal pads, including the most distal arolium, which is used for generating adhesion. (B) A tethered insect on the frustrated total internal reflection (FTIR) experimental setup (see Figure \ref{fig:methods01} for details). (C) The view of the insect through the FTIR setup, where the arolia reflect the trapped light when in contact with the glass substrate. Scale bars represent: (A-C) 20 mm and (i) 1 mm.}
\label{fig:intro1}
\end{figure}

In order to stick to natural surfaces, certain insects, like the Indian stick insect (\textit{Carausius morosus}) shown in Figure \ref{fig:intro1}A, have developed smooth and wet adhesive pads on their feet, which are unlike the hairy and dry adhesive pads observed on the toes of geckos \cite{autumn-2008-philtransrsoca}. To facilitate wet adhesion, smooth adhesive pads secrete an adhesive liquid into the contact zone between the pad and the substrate \cite{dirks-2011-softmatter}. The contact is mediated by a thin film of this adhesive liquid, which increases the pad's effective contact area \cite{dirks-2011-softmatter}.

The typical models of wet adhesion of insects consider two undeformable flat substrates, separated by a continuous liquid layer \cite{walker-1993-intjadhesadhes, federle-2002-integrcompbiol}. A liquid bridge is formed between the two surfaces, and the total adhesive force is simply given by the sum of the surface tension, Laplace pressure, and viscous Stefan adhesion \cite{hanna-1991-jexpbiol}. However, the major drawback of such a system lies in the low adhesive strengths ($\sim$ 1 MPa) that can be achieved as compared to dry adhesion ($\sim$ 20 MPa) \cite{smith-1991-jexpbiol}. The difference can be overcome by making the adhesive pads deformable \cite{dirks-2014-beilsteinjnanotechnol}. 

Insects and tree frogs have been observed to have smooth and soft adhesive pads \cite{federle-2006-jrsocinterface}, with a sponge-like structure \cite{book-gorb}. A soft adhesive pad (with a low elastic modulus) deforms more easily at a given external force, resulting in a larger contact area \mbox{\cite{buscher-2021-beilstein}}. This higher contact area in turn increases the contact radius of the mediating liquid as the liquid is pressed towards the outside of the pad, which then increases the capillary force \cite{butt-2010-softmatter}. The Young's modulus of the soft pad also plays a role in determining the capillary tension \cite{wexler-2014-prl}. However, this deformability invalidates existing adhesion models that rely on viscous Stefan adhesion, which only consider undeformable substrates \cite{book-gorb}. In the past few years, there have been a few studies \cite{duprat-2012-nature, duprat-2020-softmatter, butler-2019-prf} on the liquid-mediated adhesion between two soft elastic substrates. However, the existing models are not based on \textit{in situ} measurements of live insects, thus suggesting that there is room for improvement.

For an adhesive pad of area $A$ that must support a mass $m$, the following scaling is expected: $A \sigma \sim m$ \cite{labonte-2016-pnas}, where $\sigma$ is the adhesive strength (or stress) of the pad. However, this follows the assumption that the nature of the adhesive force acts per unit area, akin to a Laplace pressure or constant adhesive stress. Moreover, the total available area of biological adhesive pads was found to exhibit positive allometry, with the area $A$ scaling directly with the mass $m$ of the organism, or $A \sim m^1$ \cite{labonte-2016-pnas}, which implies that $\sigma \sim m^0$, i.e. biological adhesive pads generate the same adhesive strength regardless of size and species. However, for large animals, the adhesive pad area $A$ then increases disproportionately faster than the body mass $m$. As \citet{labonte-2016-pnas} pointed out, if we extrapolate this to a human, nearly half of their total surface area would need to be adhesive in order to fully support their weight, which is, of course, not desirable if one wants to scale up an adhesive system. 

On the other hand, in the same work \cite{labonte-2016-pnas}, it was found that this direct scaling between $A$ and $m$ only holds true across all animals possessing such adhesive pads, i.e., insects, arachnids, reptiles, and mammals, whereas adhesive pad area was found to scale isometrically, or $A \sim m^{2/3}$, within respective phylogenetic levels. Therefore, phylogenetic inertia (or phylogenetic constraint), or the tendency for previous adaptations to influence future adaptations \cite{mckitrick-1993-phylogenetic,darwin-1859-origin}, seems to limit how large adhesive pads can grow within a species. This issue of scaling gives rise to the following questions: (i) Do adhesive pads in whole insects exhibit the same stickiness, or adhesive stress, across size (or body mass)? (ii) Are existing mathematical models of capillary-based adhesion capable of predicting the adhesive performance of stick insects? (iii) What are the desired physical properties of the secreted liquid in order to adhere to smooth substrates?


In this paper, we address these questions through a combination of tethered experiments to measure the adhesive force of whole insects (Figure \ref{fig:intro1}B), frustrated total internal reflection (FTIR) for visualizing the contact geometry of the insects' adhesive footpads (Figure \ref{fig:intro1}C), and mathematical modelling to interpret the results and predict the physical properties of the secreted liquid in order to inform the design of biomimetic adhesives. The experiments are conducted with live Indian stick insects (\textit{Carausius morosus}) spanning their life cycle and more than two orders of magnitude in size (body mass $m$), with simultaneous and synchronized force and FTIR measurements to directly probe the relationship between adhesive force $F$, contact area $A$, adhesive stress $\sigma$, contact perimeter $P$, pad sliding distance $\delta$, and body mass $m$.

\section*{Materials and Methods}
\subsection*{Study Animals}
Female Indian stick insects (\textit{Carausius morosus}) were obtained as nymphs from Mierenboerderij (Apeldoorn, The Netherlands \url{https://www.mierenboerderij.nl/}). They were kept at 22.5 $^{\circ}$C and 50 \% relative humidity, and were fed European ivy (\textit{Hedera helix}) that was picked from around Wageningen University in Wageningen, The Netherlands.

\subsection*{Tethering Animals}
In order to tether an insect for an experiment, the insect was first sedated using CO$_2$, unless it was a fully grown adult ($m>500$ mg) and sedation wasn't necessary. For sedation, the insect was placed on a porous block and CO$_2$ was infused at a volumetric flow rate of approximately 1.0 m$^3$/h. After the stick insect was sedated, two ends of a fishing line (Nanofil size 0.04, with 0.0545-mm diameter) were fastened to both ends of the dorsal side of the thorax of the stick insect using UV-curable glue (Norland optical adhesive, type 60). The fishing line was glued between their hindlimbs and forelimbs, as depicted in Figure \ref{fig:intro1}B, to apply a pulling force equally across the limbs and prevent pitching rotation of the body. Then, their body mass $m$ was measured with a precision mass balance (Ohaus Corporation Adventurer Pro AV114CU, with 0.1 mg resolution).

\begin{figure}[bt]
\centering
\includegraphics[width=\textwidth]{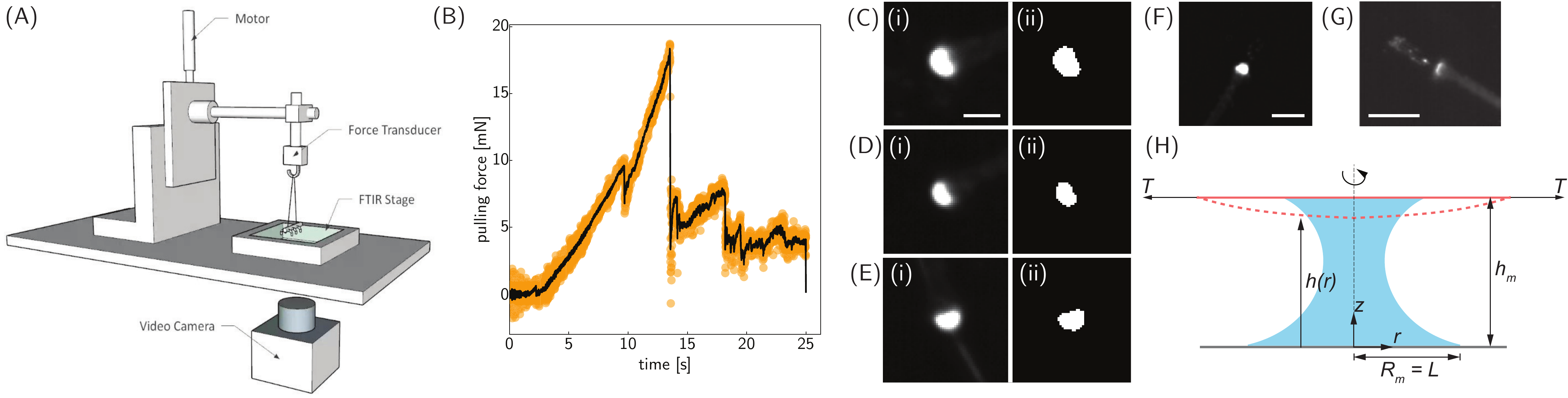}
\caption{(A) Schematic of the experimental setup combining tethered pulling measurements with frustrated total internal reflection (FTIR) imaging. (B) Pulling force measurement for a typical experiment with an insect of mass $m = 390$ mg. The discrete data points in yellow represent the experimental measurements, while the solid black line denotes the filtered data. The adhesive force $F$ is taken as the peak force minus the insect's weight. (C-E) Images of three adhesive pads, from the same insect as in panel B, using FTIR with (i) raw and (ii) binarized images. The binarized images are used for quantifying the contact area $A$ and the contact perimeter $P$. Scale bar represents 1 mm. (F) Image of a footpad after sliding, from the same insect as in panels B-E. Scale bar represents 2 mm. (G) Image of a footpad after sliding, from an insect with $m=21$ mg. Scale bar represents 1 mm. (H) Schematic for mathematical modeling, inspired by \citet{butler-2019-prf}. Here, the solid pink line denotes the undeformed footpad while the dashed line represents the deformed footpad. The liquid bridge is represented in blue while the grey line denotes the substrate to which the insect is adhering.}
\label{fig:methods01}
\end{figure}

\subsection*{Force Measurements and Frustrated Total Internal Reflection (FTIR)}
A tethered and awakened stick insect was positioned on the borosilicate glass plate in the test setup (Figure \ref{fig:methods01}A). The glass plate had LED strips along each side and was mounted on a table with a rectangular hole. A high-speed camera (Microtron CXP25) was mounted underneath the table to take recordings of the contact geometry of the stick insects. For the smaller stick insects ($m<50$ mg), a Nikon AF Micro Nikkor 105 mm f/2.8 lens was used, with a spatial resolution of 19 $\mu$m px$^{-1}$, while, for the larger stick insects ($m>50$ mg), a Nikon AF Nikkor 50 mm f/1.8D lens was used, with a spatial resolution of 44 $\mu$m px$^{-1}$.

MATLAB (R2015b) was used to control a linear motor (Thorlabs, Z825B) that pulled the tethered insect across a distance of 25 mm vertically upward from the glass plate. A 3D-printed hook was affixed at the end of the tether to pull onto the stick insect via the tethering wire. Both the high-speed camera and the tether were synchronized by starting the filming, force recording, and pulling procedure at the same time. Each experiment lasted 25 seconds, with a pulling speed of 1 mm s$^{-1}$. The experiments were conducted at $21.0-25.8$ $^{\circ}$C and $26-61$ \% relative humidity.

\begin{figure}[h]
\centering
\includegraphics[width=0.5\textwidth]{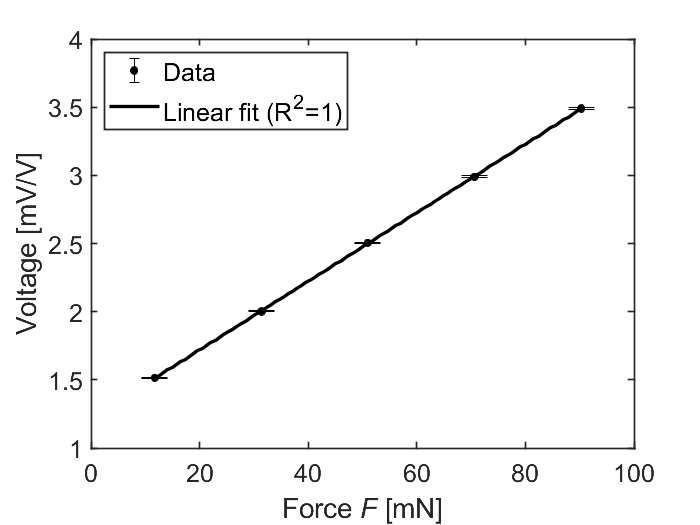}
\caption{Calibration of the force sensor. The error bars represent standard deviation from 10 second measurements.}
\label{fig:calib}
\end{figure}

\begin{figure}[h]
\centering
\includegraphics[width=\textwidth]{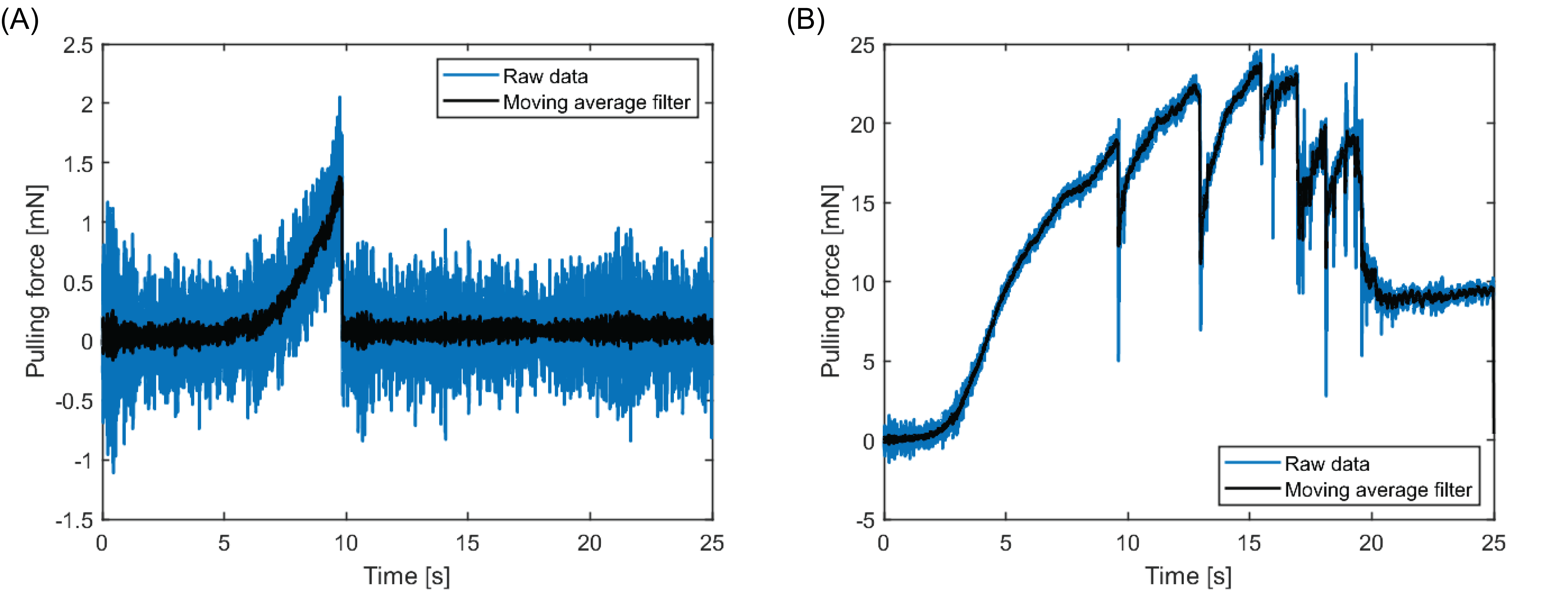}
\caption{Pulling force measurements for experiments with insects of mass (A) $m = 4.9$ mg and (B) $m = 1200$ mg. The raw data are shown in blue and filtered data in black. For (A), the peak force occurred at a time of 9.7 s, while, for (B), the peak force occurred at a time of 15.4 s.}
\label{fig:pull}
\end{figure}

A force transducer (Futex LSB200, 10g capacity) measured the force pulling on the insect at a sampling rate of 1000 Hz. Before the experiment, the force transducer was calibrated with 4 different weights. The slope of the linear regression line through the 4 data points was used to calculate force (in mN) from the measured change in electrical voltage (in mV V$^{-1}$) by the transducer. Figure \mbox{\ref{fig:calib}} shows the calibration data and the linear regression. The measured pulling force for each insect was filtered using a moving average filter with a window size of 25, and the peak (or maximum) force was extracted. Following a free-body diagram, the insect's weight $m g$ was subtracted from the peak force to obtain the adhesive force $F$. The temporal variation of the pulling force from a typical experiment is shown in Figure \ref{fig:methods01}B. Figure \mbox{\ref{fig:pull}}A-B show two other examples from the smallest ($m = 4.9$ mg) and largest ($m = 1200$ mg) insects used in the study.

The contact area was captured by the high-speed camera at 124 Hz with 5 MPx resolution using frustrated total internal reflection (FTIR). A light beam will mostly reflect internally when it is shone into a medium that has a higher index of reflection compared to the air surrounding it. However, when another object with a similar index of refraction comes into contact with the medium, then some of the total internal reflection will be frustrated and scatter out of the medium. FTIR was used to visualize the contact geometry of the adhesive pads on the glass plate, so that the contact area $A$ and perimeter $P$ could be quantified.

Using the filtered force measurement, the time of peak force was obtained and the associated image was analyzed in MATLAB (R2018b) to obtain the contact area $A$ and contact perimeter $P$ at the point of peak force generation. In order to quantify these parameters, the image was binarized using a threshold of 0.5. Then, the center of each pad was identified by manual clicking, and a square region of 30 by 30 pixels (enough to encompass each entire pad) was drawn around each pad. Typical images obtained using the FTIR method are shown in Figure \ref{fig:methods01}C-E, with the raw images on the left (i) and binarized images on the right (ii). 

To obtain the contact area $A$, the total number of pixels within each square was summed and combined with a calibration value to obtain the contact area of each pad. The area of the pads from all the limbs of the insect was summed to determine the total contact area $A$. Finally, to obtain the contact perimeter $P$, the $bwperim$ function in MATLAB (R2018b) was used to calculate the perimeter of each pad and then summed.

Using both force and contact geometry measurements, the adhesive stress $\sigma$ was determined using the following expression:

\begin{equation}
\sigma = \frac{F}{A} \, .
\label{eq:stress}
\end{equation}

\subsection*{Power law fitting}
After the adhesive force $F$, contact area $A$, contact perimeter $P$, and adhesive stress $\sigma$ were determined for each insect of mass $m$, the data were log transformed in order to determine the best-fitting power law. The log-transformed data were fit using a linear regression, with the slope of the regression representing the exponent of the best fitting power law. The data were plotted in log-log axes along with the best power-law fits in Figure \ref{fig:results1}A-D, and the power-law exponents with 95\% confidence interval (CI) and coefficient of determination R$^2$ are shown in Table \ref{tab:results1}.

\subsection*{Measurements of footpad sliding}
Using the synchronized high-speed videos from the FTIR set-up, we quantified the sliding distance $\delta$ of the footpads of the stick insects. After finding the video frame associated with the point of maximum adhesive force $F$, the frame when the footpads started to slide was found for each trial. Then, using the $imshowpair$ function in MATLAB (R2018b), the two images (at the onset of sliding and when maximum force occurred) were overlaid and the sliding distance of each footpad was measured by clicking the center of each footpad before and after sliding. Finally, the sliding distances of the 6 footpads were averaged for an individual and reported as the sliding distance $\delta$ (Figure \mbox{\ref{fig:results1}}F). Figure \mbox{\ref{fig:methods01}}F-G show images of footpads after sliding.

\subsection*{Mathematical Model}

The stick insects studied in the current work have smooth, deformable footpads that secrete a liquid. Thus, they create a liquid bridge between the pad and the substrate, and make use of the elastocapillarity arising from both the elasticity of the deformable footpad and the capillarity of the liquid bridge to adhere. The fluid dynamics of this particular scenario has been recently studied by \citet{butler-2019-prf}. In what follows, we briefly describe the model used in our study, which is largely based on the model proposed by \mbox{\citet{butler-2019-prf}}, and the assumptions and modifications we made.

The schematic of the model geometry is shown in Figure \ref{fig:methods01}H. We consider the insect footpad to be a circular, deformable membrane of constant thickness and having a Young's modulus $E$. We restrict ourselves to small axisymmetric deformations of the membrane about the coordinate system shown in Figure \ref{fig:methods01}H. We assume that the imposed tension $T$ on the footpad by the insect is uniform, and that the ends of the membrane (footpad) are fixed at the radial position $r = L$. Hence, by modulating the tension $T$, the insect can only change the curvature of the membrane. Note that in reality $T$ can change due to the vertical deformation of the membrane, but we neglect that here for simplicity (also following \citet{butler-2019-prf}). The insect adheres to a flat, smooth, and rigid substrate at a vertical distance $h_{M}$ from the membrane by secreting a liquid of volume $V$ and surface tension $\gamma$ between the membrane and the substrate. From previous experiments \cite{federle-2002-integrcompbiol}, we know that the distance between the membrane and the substrate is small, which means that the aspect ratio $h_{M} / L \ll 1$. Note that we exaggerate this gap in Figure \ref{fig:methods01}H for clarity. Previous studies \citep{dirks-2010-jrsocinterface} suggest that the secreted liquid volume $V$ is small; hence gravity can be neglected in the mathematical formulation. 

The interfacial tension $\gamma$ of the liquid bridge formed between the membrane and the substrate results in a capillary force due to the pressure difference between the inside and the outside of the liquid volume. This pressure difference is proportional to the curvature of the liquid bridge. We further assume that the liquid secretion perfectly wets the footpad and the substrate, as supported by previous studies that found the liquid secreted by the pads is also secreted throughout the rest of the body \cite{geiselhardt-2009-secrete}. Hence, the liquid covers the entire surface area of the membrane (note that Figure \ref{fig:methods01}F shows the more generalized case where the liquid only partially wets the membrane). The interfacial tension is expected to contribute to a discontinuity of the membrane tension at the membrane-liquid contact line \citep{butler-2019-prf}, but this is neglected here as $T \gg \gamma$. While smooth adhesive pads of insects are soft in compression, in order to conform to rough substrates, their internal fibrillar structure provides high resistance to tension \mbox{\cite{gorb-2007-advinsectphysiol}}.

For the squeezed thin liquid bridge ($h_{M} / L \ll 1$) in the present scenario, we can assume that the axial curvature dominates over the azimuthal curvature in determining the capillary force \citep{reyssat-2015-jfm}. We further approximate the meniscus cross-section to be a circular arc of radius $h_{M}/2$ (since the liquid is perfectly wetting the membrane and the substrate with very low contact angles \cite{federle-2002-integrcompbiol}) \citep{butler-2019-prf}. The capillary pressure at the meniscus (relative to the atmospheric pressure) can then be written as 

\begin{equation}
	p_{M} = - \frac{2 \gamma}{h_{M}} \, .
	\label{eq:cap-pressure}
\end{equation}

Before we discuss adhesion with a deformable membrane, it is worth visiting the classical limiting case of adhesion with a rigid membrane, i.e. when $T \rightarrow \infty$. The adhesive force results from the capillary pressure $p_{M}$ acting over an area $A = V / h_{M}$, and is given by

\begin{equation}
	F_{\mathrm{rigid}} = 2 \gamma \frac{V}{h_{M}^{2}} \, .
	\label{eq:force-rigid}
\end{equation}
Thus, the adhesive force in this case is purely governed by the separation gap $h_{M}$. 

Let us now consider the deformable membrane, which is the relevant scenario in the present case. For the axisymmetric coordinate system shown in Figure \ref{fig:methods01}H, the membrane position is described by $z = h(r,t)$ with the substrate at $z = 0$, where $r$ is the radial coordinate, $z$ the axial coordinate, and $t$ the time. We consider the static scenario (no flow within the liquid volume) where the pressure field $p(r,t)$ within the liquid is uniform, and the membrane shape $h(r,t)$, determined by a local force balance, is a solution of the Young-Laplace equation

\begin{equation}
	\frac{T}{r} \frac{\partial}{\partial r} \left( r \frac{\partial h}{\partial r} \right) = - p \, ,
    \label{eq:diffeq-p}
\end{equation} 
where we have assumed a small membrane slope because of the small aspect ratio ($h_{M} / L \ll 1$) and neglected the inertia of the membrane. 

The secreted liquid volume is also constant, which results in the following conservation equation:
\begin{equation}
	V = 2 \pi \int_{0}^{L} r h \, \mathrm{d}r \, ,
\end{equation}
where we have assumed that the meniscus shape has a negligible effect on the volume due to the small aspect ratio ($h_{M} \ll L$).

Since we consider here the equilibrium scenario where there is no flow within the liquid volume, the pressure within the liquid, $p$, must be uniform and equal to the pressure at the meniscus, $p_{M}$ (given by equation \eqref{eq:cap-pressure}). Hence, equation \eqref{eq:diffeq-p} can be rewritten as
\begin{equation}
    \frac{1}{r} \frac{\partial}{\partial r} \left( r \frac{\partial h}{\partial r} \right) = \frac{2 \gamma}{T h_{M}} \, ,
    \label{eq:diffeq-h}
\end{equation}
subject to the boundary conditions arising from the imposed symmetry at $r = 0$ and the meniscus position at $r = r_{M}$ (note that here $r_{M} = L$ since we have assumed that the liquid perfectly wets the membrane):
\begin{equation}
    \frac{\mathrm{d}h}{\mathrm{d}r} =0 \,\,\,\,\, \mathrm{at} \,\,\,\,\, r = 0 \,\\,\,\,\, \mathrm{and}
    \label{eq:bc-1}
\end{equation}
\begin{equation}
    h = h_{M} \,\,\,\,\, \mathrm{at} \,\,\,\,\, r = r_{M} \, .
    \label{eq:bc-2}
\end{equation}
The governing equation \eqref{eq:diffeq-h}, along with the boundary conditions \eqref{eq:bc-1} and \eqref{eq:bc-2} leads to the radial height profile given by
\begin{equation}
    h = \frac{\gamma}{T h_{M}} \left( \frac{r^{2} - r_{M}^{2}}{2} + h_{M}^{2} \right) \, .
    \label{eq:h-profile}
\end{equation}
While all of these are interesting results, perhaps the quantity most relevant to the present research is the adhesive force $F$. We restrict ourselves to the `non-contacting' scenario \citep{butler-2019-prf}, where the membrane does not touch the substrate, i.e. there is always a thin liquid layer between the membrane and the substrate. In such a case, following from equation \mbox{\eqref{eq:force-rigid}}, the adhesive force can be expressed as
\begin{equation}
    F = 2 \pi \gamma \frac{r_{M}^{2}}{h_{M}} \, ,
    \label{eq:force}
\end{equation}
which indicates that the adhesive force scales with the square of the contact radius but inversely with the liquid film thickness. In what follows, we use this result to rationalize our experimental findings and discuss their implications. 

\section*{Results}
\begin{figure}[bt]
\centering
\includegraphics[width=\textwidth]{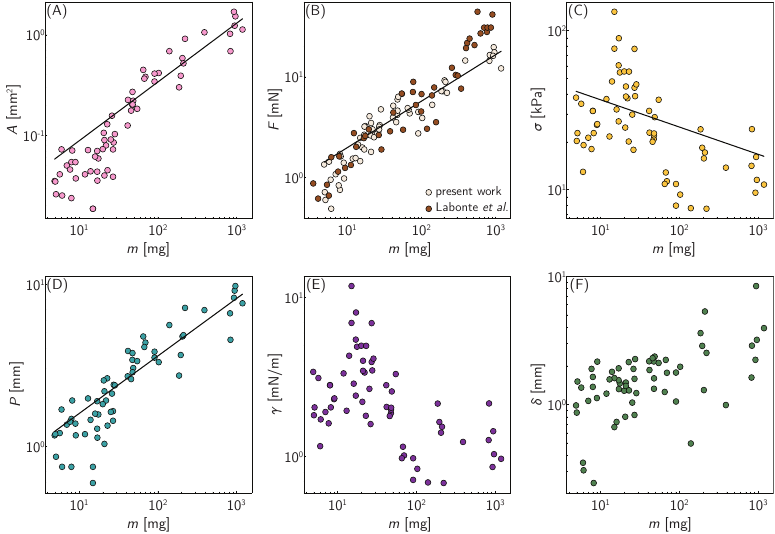}
\caption{Measurements of: (A) contact area $A$, (B) adhesive force $F$, (C) adhesive stress $\sigma$, (D) contact perimeter $P$, (E) liquid tension $\gamma$, and (F) sliding distance $\delta$ across body mass $m$. The solid lines denote power-law fits, provided in Table 1. Panel B also shows the whole-insect, centrifuge measurements from \mbox{\citet{labonte-2019-procrsocb}}}.
\label{fig:results1}
\end{figure}

By combining the tethered force measurements with the FTIR imaging, the adhesive force and pad contact geometry were measured simultaneously for $N=63$ Indian stick insects varying in body mass $m$ from 4.9 to 1200 mg. The results are shown in Figure \ref{fig:results1}, with the power-law fittings provided in Table \ref{tab:results1}.

For contact area $A$ (Figure \ref{fig:results1}A), it was found to scale as $m^{0.77}$ (95\% CI: 0.69, 0.85 and R$^2 = 0.86$), while adhesive force $F$ (Figure \ref{fig:results1}B) was found to scale as $m^{0.55}$ (95\% CI: 0.49, 0.60 and R$^2 = 0.87$). The combination of these two scalings, via equation \eqref{eq:stress}, reflects what was found for adhesive stress $\sigma$ (Figure \ref{fig:results1}C), which scaled as $m^{-0.22}$ (95\% CI: -0.31, -0.13 and R$^2=0.27$). Therefore, the adhesive strength of the pads decreased as the insects grew in size. A Spearman's rank correlation indicated that stress $\sigma$ and mass $m$ were correlated with a decreasing trend ($\rho=-0.52$ and $p=1.4 \times 10^{-6}$). The contact perimeter $P$ (Figure \ref{fig:results1}D) was found to scale as $m^{0.41}$ (95\% CI: 0.35, 0.46 and R$^2=0.79$).



Using the mathematical model (equation \mbox{\eqref{eq:force}}), the surface tension $\gamma$ of the adhesive liquid secretion was predicted (Figure \mbox{\ref{fig:results1}}E). The height $h_M$ of the liquid layer was assumed to be 90 nm, as previously measured using interferometry \mbox{\cite{federle-2002-integrcompbiol}}. The predicted surface tension $\gamma$ ranged between 0.68 mN m$^{-1}$ and 12 mN m$^{-1}$, which is lower than the surface tension for oil-based liquids of approximately 20 mN m$^{-1}$, which is the value typically assumed for the secreted liquid \mbox{\cite{gernay-2016-interface,gennes-2004-capillarity}}. If we instead assume surface tension $\gamma = 20$ mN m$^{-1}$, then the predictions of liquid height $h_M$ range from 150 nm to 2600 nm, which are greater than those measured using interferometry (90 nm to 160 nm) \mbox{\cite{federle-2002-integrcompbiol}}.

Figure \mbox{\ref{fig:results1}}F shows how sliding distance $\delta$, which is associated with shearing of the footpads, varies with body mass $m$. In the live insects, we observed a similar relationship as was reported for the controlled, single-pad measurements from \mbox{\citet{labonte-2019-procrsocb}}. The sliding distance $\delta$ was relatively constant for small insects, and then increased for larger insects. Using a Spearman's rank correlation, we found that sliding distance $\delta$ and mass $m$ were correlated with an increasing trend ($\rho=0.52$ and $p=4.5 \times 10^{-6}$).

\begin{table}[bt]
\caption{Summary of the power-law fits for measurements shown in Figure \ref{fig:results1}A-D.}
\begin{threeparttable}
\begin{tabular}{lccrr}
\headrow
\thead{Parameter} & \thead{Unit} & \thead{Power-law exponent} & \thead{95\% CI} & \thead{R$^2$}\\
Contact area $A$ & mm$^2$ & 0.77 & (0.69, 0.85) & 0.86\\
Adhesive force $F$ & mN & 0.55 & (0.49, 0.60) & 0.87\\
Adhesive stress $\sigma$ & kPa & -0.22 & (-0.31, -0.13) & 0.27\\
Contact perimeter $P$ & mm & 0.41 & (0.35, 0.46) & 0.79\\
\hline  
\end{tabular}
\begin{tablenotes}
\item CI, confidence interval; R$^2$, coefficient of determination.
\end{tablenotes}
\end{threeparttable}
\label{tab:results1}
\end{table}


\section*{Discussion}
The adhesive pad area was previously measured for various animals using microscopic images \cite{labonte-2016-pnas}. For Indian stick insects, the area was found to scale as $m^{0.70}$, following what is expected from isometry, with area scaling as $m^{2/3}$. The measurements reported here show a scaling slightly higher than isometry, but with the lower bound of the 95\% confidence interval overlapping with the previous pad area measurements of Indian stick insects from \mbox{\citet{labonte-2016-pnas}}. In addition to quantifying contact area $A$, we measured the contact perimeter $P$ of the pads and find that the perimeter $P \sim A^{1/2}$, as expected from isometry.



Previous research investigated the scaling of Indian stick insect adhesion by combining whole-insect experiments with a centrifuge and single-pad measurements with a motorized stage and feedback loop \cite{labonte-2019-procrsocb}. In this work, it was found that the adhesive force $F \sim m^{0.69}$ (95\% CI: $0.59–0.79$) in whole insects. For single pads, $F \sim m^{0.34}$ (95\% CI: $0.27–0.40$) without shearing, $F \sim m^{0.71}$ (95\% CI: $0.61–0.82$) with shear force proportional to $m^{2/3}$, and $F \sim m^{0.87}$ (95\% CI: $0.70–1.03$) with shear force proportional to $m^{1}$. While the forces measured here are similar to those reported for the centrifuge measurements (see Figure \mbox{\ref{fig:results1}}B), the scaling with mass differs slightly, with a small overlap of the 95\% confidence intervals. 



In our direct measurements, we observed that stress $\sigma$ decreased with body mass $m$; however, with a poor goodness of fit (R$^2=0.27$). Therefore, it is possible that adhesive stress $\sigma$ is instead independent of body mass $m$. According to \mbox{\citet{labonte-2019-procrsocb}}, if the stress $\sigma$ decreases with mass $m$, or $\sigma \sim m^{-1/3}$, it would indicate that the insects are not shearing their adhesive pads. However, as shown in Figures \mbox{\ref{fig:methods01}}F-G and \mbox{\ref{fig:results1}}F, the insect pads were observed to shear, with the sliding distance $\delta$ varying between 0.24 mm and 8.4 mm, which is significantly higher than the range of sliding distances reported for the single-pad measurements in \mbox{\cite{labonte-2019-procrsocb}} (0 mm to 2 mm). Similarly, a previous study on beetles \mbox{\cite{amador-2017-jroysocint}} observed greater sliding distances in live, freely climbing beetles when compared to controlled single-pad experiments with simulated steps \mbox{\cite{clemente-2010-jexpbiol}}.

Previous mathematical models have been proposed for capillary-based adhesion. However, many of these don't account for the height $h_M$ of the liquid film. For instance, the capillary adhesion model used to predict the attachment performance of an array of small liquid bridges inspired by a leaf beetle states that $\sigma \sim P^{-1}$ \mbox{\cite{vogel-2010-pnas}}, neglecting height altogether. While we also found a decreasing trend in adhesive stress $\sigma$ versus size, we don't find the same inverse scaling relation between stress $\sigma$ and perimeter $P$. Another capillary adhesion model, based on Hertz contact theory of elastic solids with attraction effects, via the extension by Johnson, Kendall, and Roberts (JKR theory), predicts that $F \sim R_c$, where $R_c$ is the radius of curvature of the adhesive pad \mbox{\cite{federle-2002-integrcompbiol,fogden-1990-contact}}. For our measurements, we find that adhesive force $F$ does not seem to scale with pad radius (or $A^{1/2})$. However, it remains unknown whether radius of curvature $R_c$ scales isometrically with body mass $m$.

For hairy adhesive pads with secreted liquid, like in the green dock beetle (\textit{Gastrophysa viridula}), the elastocapillary adhesion of individual fibers was modeled using a similar capillary model as in equation \mbox{\eqref{eq:force}}, based on the capillary Laplace pressure \mbox{\cite{gernay-2016-interface}}. Using this model, the adhesive force that each fiber generated was predicted and found to agree with previous experiments on single fibers of the same species \mbox{\cite{gernay-2016-interface,bullock-2011-beetle}}.

From our predictions, based on equation \mbox{\eqref{eq:force}}, we find that if the liquid height $h_M$ is constant, then the secreted adhesive liquid does not require high surface tension in order to generate the observed adhesive stress $\sigma$. An average surface tension of $\gamma=2.7$ mN m$^{-1}$ was found to be sufficient, given a liquid film height $h_M = 90$ nm. Therefore, it is possible that stick insects prioritize the wettability of their secreted adhesive liquid. Secreting a liquid with such a low surface tension may explain how versatile insect adhesion is with respect to substrate properties. A previous study with three stick insect species found that adhesive force was not significantly affected by the surface free energy of the substrate \mbox{\cite{thomas-2023-jeb}}. With low surface tension, the secreted liquid can easily flow into the asperities on rough substrates in order to maximize contact area. By using the capillary model proposed by \mbox{\citet{butler-2019-prf}} (equation \mbox{\eqref{eq:force}}), engineers can make informed decisions on the development of capillary-based adhesives, especially regarding the working liquid.

The capillary model we used, from \mbox{\citet{butler-2019-prf}}, assumes a static situation, where the pad is not sliding and the secreted liquid is not flowing. We did observe significant sliding in our experiments (see Figure \mbox{\ref{fig:results1}}F), so future work should be dedicated to developing a dynamic model that includes the effects from shearing. Previous work has shown that adhesive force $F$ is linearly proportional to shear force \mbox{\cite{labonte-2019-procrsocb,federle-2019-procrsocb}}; however, the mechanisms underlying this linear relationship are still unknown. Therefore, shearing should be accounted for in a more sophisticated model in order to determine if the liquid height $h_M$ is affected by the amount of shearing and how this could relate to the size of the insect, especially since the sliding distance $\delta$ was observed to increase with mass $m$ (Figure \mbox{\ref{fig:results1}}F) and the secreted liquid is deposited onto the substrate during shearing (see Figure \mbox{\ref{fig:methods01}}F-G and \mbox{\citet{labonte-2019-procrsocb}}). Moreover, the inverse relationship between adhesive force $F$ and liquid height $h_M$ has been found to not hold true when attaching to rough substrates \mbox{\cite{dirks-2014-beilsteinjnanotechnol,baier-1968-science,drechsler-2006-jcompphysiola}}. Therefore, further developments in mathematical modeling should also aim to include the effects of substrate roughness on adhesion force.

\section*{Acknowledgements}
We thank Remco Pieters and Anne de Waal for assistance with the experimental setup, Jelle Steens and Susan van den Bos for their early contributions in designing the experiments, and David Labonte for the fruitful discussions and review of the manuscript.

\section*{Conflict of interest}
The authors declare no conflicts of interest.

\section*{Supporting Information}

\printendnotes

\newpage
\bibliography{bioadhesion}

\end{document}